\documentclass{PoS}
\usepackage{amsmath}
\usepackage{mathtools}
\usepackage{floatrow}
\def\bracketbar{\hbox{\kern-8pt\raise1pt%
    \hbox{{\tiny(}{\lower1.5pt\hbox{\bf--}}{\tiny)}}}}

\let\oldbibliography\thebibliography
\renewcommand{\thebibliography}[1]{%
  \small
  \oldbibliography{#1}%
  \setlength{\itemsep}{0pt}%
}

\title{Measurements of neutrino-nucleus scattering}

\ShortTitle{Neutrino-nucleus scattering}

\author{\speaker{C. Wilkinson}\\
        University of Bern, Albert Einstein Center for Fundamental Physics, Laboratory for High Energy Physics (LHEP), Bern, Switzerland\\
        E-mail: \email{callum.wilkinson@lhep.unibe.ch}}

\abstract{Current and planned neutrino oscillation experiments operate in the 0.1-10 GeV energy regime. At these energies, the neutrino cross section is not well understood: a variety of interaction processes are possible and nuclear effects play a significant role. Here, the conceptual problems that affect measuring and understanding neutrino cross sections are introduced, and the status of neutrino cross section measurements for CC0$\pi$ and CC1$\pi$ channels are discussed.}

\FullConference{Neutrino Oscillation Workshop (NOW2018)\\
		9--16 September, 2018\\
		Rosa Marina (Ostuni, Brindisi, Italy)}

\begin{document}

\section{Introduction}
\begin{figure}[htbp]
  \centering
  \floatbox[{\capbeside\thisfloatsetup{capbesideposition={right,center},capbesidewidth=0.45\textwidth}}]{figure}[\FBwidth]
         {\caption[fragile]{Neutrino fluxes, as a function of energy, of current and future accelerator-based neutrino experiments. The fluxes relevant to neutrino interaction measurements are shown, from the T2K experiment off-axis near detector (ND)~\cite{Abe:2012av}, NOvA ND~\cite{novaflux}, and MINERvA (Low Energy configuration only~\cite{Aliaga:2016oaz}). One future program, DUNE~\cite{Acciarri:2015uup}, is shown; the HK flux is very similar to the T2K ND off-axis flux. Also overlaid are the total CC cross section divided by energy, according to GENIE 2.12.8~\cite{Andreopoulos:2009rq}. Figure reproduced from Ref.~\cite{Mahn:2018mai}.}
  \label{fig:flux_and_xsec}}
         {\includegraphics[width=0.5\textwidth]{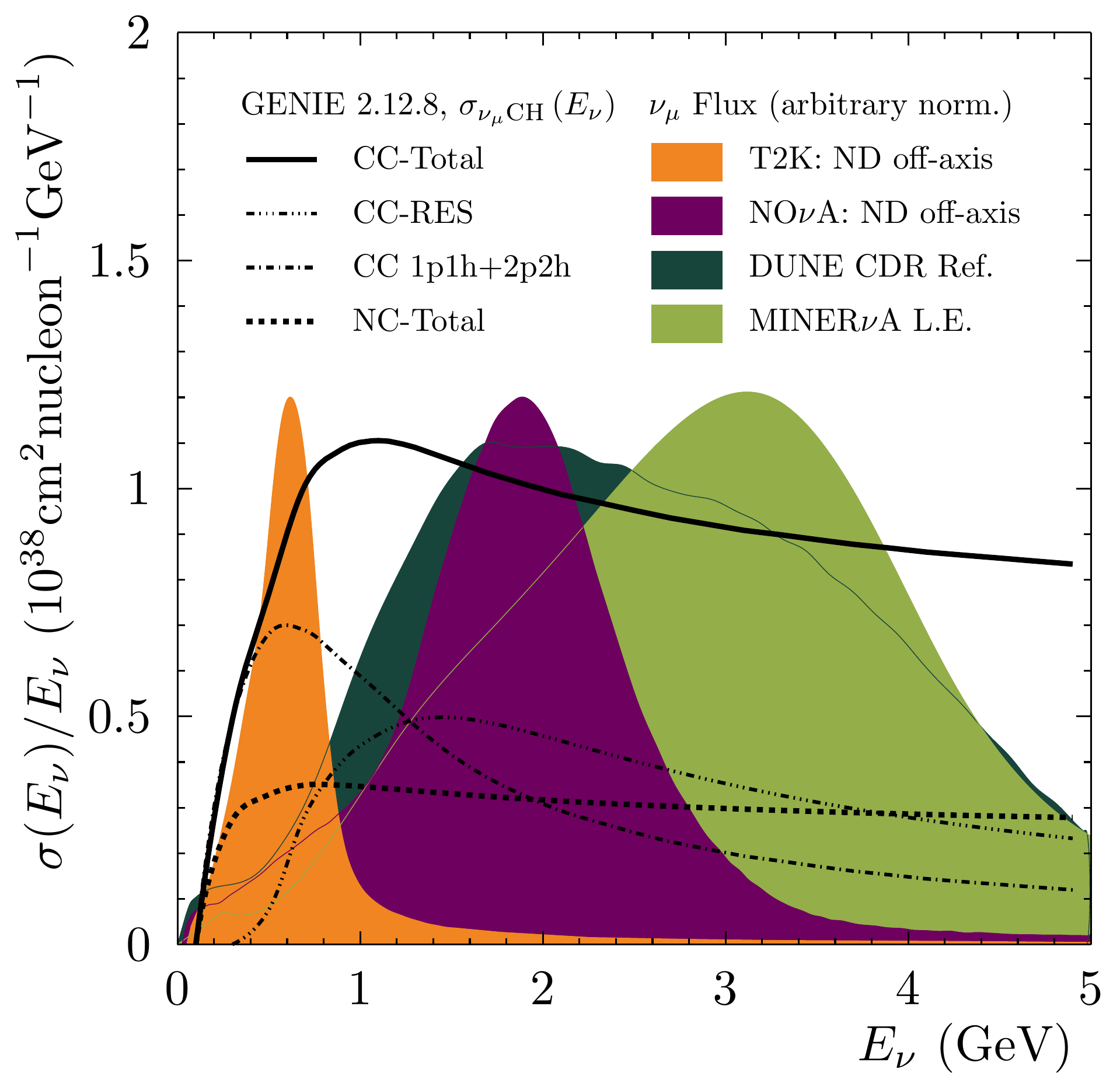}}
\end{figure}

\vspace{-10pt}
Neutrino oscillations are a well-established phenomenon, which depend on the neutrino energy, $E_{\nu}$, and distance travelled. A number of current and planned accelerator neutrino experiments aim to make precise measurements of the underlying parameters which govern the oscillation probability. These all operate in the few-GeV energy region, in the so-called ``transition region'', where multiple neutrino interaction modes contribute, as shown in Figure~\ref{fig:flux_and_xsec}.

The basic method for conducting an accelerator neutrino oscillation analysis is to compare the rates at a near detector, located close to the production point before oscillations can occur, and at a far detector, located near the oscillation maximum. The rate in each detector, $R(\vec{\mathbf{x}})$, as a function of reconstructed final-state particle kinematic quantities, $\vec{\mathbf{x}}$, can be expressed as:
\begin{equation}
  R(\vec{\mathbf{x}}) = \underbracket[1pt][10pt]{\overbracket[1pt][10pt]{\int_{E_{\mathrm{min}}}^{E_{\mathrm{max}}}\Phi(E_\nu) \times \sigma(E_{\nu}, \vec{\mathbf{x}}) \times \epsilon(\vec{\mathbf{x}})}^{\mathrm{Near\;detector}} \times P(\nu_{A} \rightarrow \nu_{B})}_{\mathrm{Far\;detector}},
  \label{eq:rate}
\end{equation}
\noindent where $\Phi(E_\nu)$ is the neutrino flux as a function of $E_{\nu}$, $\sigma(E_{\nu}, \vec{\mathbf{x}})$ is the cross section, $\epsilon(\vec{\mathbf{x}})$ is the detector efficiency, and $P(\nu_{A} \rightarrow \nu_{B})$ is the oscillation probability. It is clear that a good understanding of the neutrino cross section, $\sigma(E_{\nu}, \vec{\mathbf{x}})$, is critical for carrying out an oscillation analysis, as it relates the neutrino energy with the variables that can be measured in the detector. Although the near detector is able to make {\it in situ} measurements of the flux, cross section and detector efficiency, which constrain uncertainties before convolving the oscillation probability at the far detector, simply taking a near/far ratio does not fully cancel systematics because there is a dramatic change in the neutrino energy and flavor distribution over which the integral runs between the detectors.

\vspace{-5pt}\section{Neutrino cross section concepts}\vspace{-10pt}
\label{sec:xsec}
A breakdown of the different major contributions to the neutrino-nucleus cross-section in the few-GeV region is indicated in Figure~\ref{fig:flux_and_xsec}. The contributions, or modes, can be conceptually separated in terms of the energy transfer to the struck nucleus. At low energy transfers, the interaction is with the nucleus as a whole, either elastically, or through the excitation of a giant nuclear resonance. At intermediate energy transfers, the interaction is with the nucleon, first as quasi-elastic scattering, or single nucleon knock-out (1p1h), $\nu^{\bracketbar}_{l} + n(p) \rightarrow l^{-(+)} + p(n)$, and then at higher energy transfers, through the excitation of a nucleon resonance (RES), which decays to produce a pion and a nucleon\footnote{Note that in rare cases, nucleon resonances can decay to final states which include multiple pions, or heavier mesons, such as kaons~\cite{pdg_2018}.}. At large energy transfers, the interaction is with a constituent parton inside the nucleon, in deep inelastic scattering (DIS). Because the nucleons are bound within the nucleus, when the interaction is with a nucleon (1p1h, RES), the response depends on the details of the initial nuclear state, in particular the initial state nucleon momentum distribution, and the energy required to liberate a bound nucleon. Additionally, interactions with more than one nucleon are possible, generally referred to as multinucleon knock-out (2p2h). The transition between interactions with an entire nucleus, to interactions with a nucleon, to interactions with a parton, is a significant challenge to building a consistent cross-section model.

Critically, because the neutrino energy and energy transfer to the nucleus cannot be reconstructed on an event-by-event basis, it is necessary to have a good understanding of the entire nuclear response, as experiments have to implicitly integrate over the energy transfer when measuring an event rate (Equation~\ref{eq:rate}). Worse, because neutrinos can interact anywhere inside the nucleus, hadrons are produced deep inside the nuclear medium, and have a high probability of interacting before they escape, and are visible in the detector. These interactions, which modify the final-state particle content, are referred to as final state interactions (FSI). As a result, we cannot measure interaction {\it modes} such as 1p1h, 2p2h or RES interactions unambiguously. Instead, we measure final state {\it topologies}, which are defined by the observable final-state particle content. For example, CC0$\pi$, where a single muon is observed, and no pions (or, generally, any other mesons). CC0$\pi$ may be dominated by 1p1h and 2p2h processes, but RES processes may also contribute if the pion produced at the vertex is absorbed through FSI.

Measurements of interaction {\it topologies} do not correct for FSI, and are therefore more model-independent than measurements of interaction {\it modes} where the experiment has imposed assumptions about FSI. However, it is a significant challenge to use measurements of interaction {\it topologies} to constrain the underlying cross-section model parameters and reduce the systematic uncertainties for oscillation experiments~\cite{Stowell:2016jfr}. Different components of the full neutrino-nucleus scattering model can be constrained through other sources to make the problem more tractable. Electron-nucleus scattering data can be used to constrain nuclear model uncertainties, as the nucleus is the same in both cases, and pion-nucleus scattering data can be used to constrain the effects of FSI to some extent (see, for example, Ref.~\cite{PinzonGuerra:2018rju}).

\vspace{-5pt}\section{CC0$\pi$ status}\vspace{-10pt}
Much of the theoretical work on neutrino-nucleus cross-section models over the last 10 years has been motivated by observed discrepancies between old models and MiniBooNE CC0$\pi$ data~\cite{mbCCQE,mbAntiCCQE}. The introduction of nuclear effects such as 2p2h, and a more sophisticated treatment of the initial nuclear state has been successful in qualitatively describing that data, and agrees with electron scattering data. However, attempts to confront the models with all available data have shown that some issues remain~\cite{Wilkinson:2015ypa, Wilkinson:2016wmz, Betancourt:2018bpu}.

There have been a number of recent measurements of CC0$\pi$ which aim to test the new theoretical models, and offer new data to help refine them. Both T2K~\cite{Abe:2016tmq, Abe:2017rfw} and MINERvA~\cite{Wolcott:2015hda, Patrick:2018gvi, Ruterbories:2018gub} have made measurements as a function of outgoing muon kinematics which show broad agreement with the available CC0$\pi$ models, but are not very sensitive to nuclear effects, and lack power to differentiate between competing models, although certain tricks can be used to improve the sensitivity~\cite{Wilkinson:2016cnr}. Additional variables, based on momentum imbalances have been proposed~\cite{Lu:2015tcr}, and measured, again by both T2K~\cite{Abe:2018pwo} and MINERvA~\cite{Lu:2018stk}. These measurements also require the reconstruction of a proton in the final state, and appear to be a powerful new tool for differentiating between models~\cite{Dolan:2018zye}. However, it is likely that more theoretical work will be needed to take full advantage of these new datasets.

Additionally, CC-inclusive datasets from MINERvA~\cite{Rodrigues:2015hik, Gran:2018fxa} which focus on low-momentum transfer kinematics offer a new probe of the region where multi-nucleon, and other nuclear effects, have a strong effect.

\vspace{-5pt}\section{CC1$\pi$ status}\vspace{-10pt}
CC1$\pi$ cross sections are more difficult to model than CC0$\pi$ cross sections as they contain significant contributions from different energy transfer regimes: coherent pion production, RES, and DIS can all produce a muon and a single pion in the final state\footnote{Measurements of coherent pion production have been carried out, and can be used to independently constrain that part of the model~\cite{AguilarArevalo:2008xs, minerva_coh_2014, Adamson:2016hyz, Abe:2016fic, Mislivec:2017qfz}.}. They are also more difficult to measure, as they contain higher multiplicity events, making a model-independent cross section more difficult to extract~\cite{Mahn:2018mai}. Attempts to compare cross-section models to CC1$\pi$ neutrino-nucleon data have shown that a consistent nucleon-level model is possible~\cite{anl_bnl_reanalysis, Rodrigues:2016xjj, Kabirnezhad:2017jmf}, but tension has been observed between the neutrino-nucleon data and neutrino-nucleus data~\cite{Stowell:2019zsh}.

A comparison between all recent neutrino-nucleus CC1$\pi$ data is given in Ref.~\cite{Mahn:2018mai}, which shows that the measured muon kinematic variables broadly agree with a reference GENIE v2.12.8 model~\cite{Andreopoulos:2009rq}, but that the measured pion kinematic variables do not, although there is some consistency between diverse datasets. This lack of overall agreement is representative of all generator models currently available. The currently available neutrino-nucleus pion-production models lack predictive power, a potentially serious issue for oscillation measurements.

\vspace{-5pt}\section{Concluding remarks}\vspace{-10pt}
Measuring neutrino cross sections, and then using those measurements to constrain a full cross section model suitable for accelerator neutrino oscillation experiments, are both significant challenges. There has been a theoretical focus on CC0$\pi$ cross section modelling over the last 10 years, which has been complemented with an experimental program to test those models. Recent, more stringent probes from the experimental community (measurements of transverse imbalances) will motivate further work to refine the theoretical models. Overall, the relationship between experiment and theory is good news for the T2K and Hyper-K experiments, which use CC0$\pi$ as a signal process. CC1$\pi$ and higher invariant mass process, are in general less well understood theoretically, without the same level of recent theory engagement as CC0$\pi$. Similarly, the measurements of CC1$\pi$ do not agree well, and indicate tension between channels and between experiments. This is a less positive picture for NOvA and DUNE, for which CC1$\pi$ is a signal process, particularly because the improvements in CC0$\pi$ modelling and measurements took many years to achieve. A final comment, relevant for DUNE, is that much of the experimental data is on hydrocarbon, or water targets, and as such, the atomic number dependence of the cross section models we have is not well tested. This will be partially mitigated by the SBN program at Fermilab, but not for the broad range of energy transfer which will be relevant for DUNE.

\vspace{-5pt}
\small{
  \bibliography{bibliography.bib}
}
\end{document}